\documentclass[journal,onecolumn]{IEEEtran}
\usepackage{amsmath, amsthm, amsfonts}
\usepackage[english]{babel}
\usepackage[dvips]{graphicx}
\usepackage{graphicx}
\usepackage{tabularx}
\usepackage{enumerate}
\usepackage{mathtools}
\usepackage[ruled]{algorithm2e}
\usepackage{algorithmic}

\addto\captionsenglish{\renewcommand{\figurename}{Fig.}}

\theoremstyle{remark}

\usepackage{subfigure}
\usepackage{cite}

\begin{document}
\title{A Family of Constrained Adaptive filtering Algorithms Based on Logarithmic Cost}

%\author{{Vinay Chakravarthi Gogineni and Mrityunjoy Chakraborty, Senior Member, IEEE}}

\author{~Vinay~Chakravarthi~Gogineni, and~Subrahmanyam ~Mula
\thanks{The authors are with the Department of Electronics and Electrical Communication Engineering, Indian Institute of Technology, Kharagpur-721302, India (e-mail: vinaychakravarthi@ece.iitkgp.ernet.in; svmula@iitkgp.ac.in)}}

\maketitle
%\pagenumbering{gobble}
\thispagestyle{empty}

\begin{abstract}
This paper introduces a novel constraint adaptive filtering algorithm based on a relative logarithmic cost function which is termed as Constrained Least Mean Logarithmic Square (CLMLS). The proposed CLMLS algorithm elegantly adjusts the cost function based on the amount of error thereby achieves better performance compared to the conventional Constrained LMS (CLMS) algorithm. With no assumption on input, the mean square stability analysis of the proposed CLMLS algorithm is presented using the energy conservation approach. The analytical expressions for the transient and steady state MSD are derived and these analytical results are validated through extensive simulations.
\end{abstract}

\section{Introduction}
Constrained adaptive filtering algorithms \cite{Wernerb} are customized for applications like linear phase system identification, antenna array processing, spectral analysis and blind multiuser detection where the unknown parameter vector need to be estimated subjecting to a set of linear equality constraints. These deterministic linear equality constraints are helpful to design a robust system and usually constructed from the \emph{a priori} knowledge about the considered problem such as linear phase in system identification, direction of arrival in antenna array processing \cite{Wernerb, Dinizb}. In this family of algorithms, Constrained Least Mean Square (CLMS) algorithm \cite{Godara, CLMSAnal} is the most popular one because of its simple structure and robustness. Many other linearly-constrained adaptive filtering algorithms \cite{CAPA, CCG, SMCAPA, CRLS} have been proposed in literature, however they require high computational effort. 
Many variants of the conventional LMS addressing different issues of the original algorithm without significantly increasing the complexity have been suggested and analyzed extensively in literature. Of particular importance is the class of LMS algorithms with error non-linearities. The earliest among them is Least Mean Fourth (LMF) algorithm \cite{LMF}, whose cost function is error raised to the fourth power instead of the mean square error used for LMS. Though LMF outperforms LMS in certain situations, LMF suffers from stability issues \cite{LMFAnal}. In an attempt to address the above problem, Least Mean Mixed-Norm \cite{LMNF} have been designed, however, the selection of mixing parameter is difficult for practical applications. The Least Mean Logarithmic Square (LMLS) algorithm proposed in \cite{LMLS} solves this issue by intrinsically combining the LMF and LMS algorithms with no need of mixing parameter thereby achieves best trade-off between convergence rate and steady-state misadjustment. In \cite{TVLSI}, it is shown that the hardware overhead of LMLS over LMS is negligible for the achieved improvement in performance and hence LMLS can potentially replace LMS in practical applications. However, we do not find any attempts so far to extend these error non-linear concepts to the constrained adaptive filtering algorithms. In this paper, we address this gap. Inspired from the recently proposed logarithmic cost based LMLS algorithm, we propose a novel Constrained Least Mean Logarithmic Square (CLMLS) algorithm which achieves a better steady-state performance and whose complexity is almost same as CLMS.

On the other hand, in many practical applications like acoustic and network echo cancellation, underwater communication \cite{Rajib}, the system (network echo path) to be estimated is sparse in nature (i.e., impulse response contains very few active coefficients while the rest of the coefficients magnitude is close to zero). These applications motivated a flurry of research activities in the area of sparse adaptive filters in the context of system identification. Unlike conventional, sparsity unaware adaptive filters like the LMS, RLS and their various variants \cite{Haykin}, these filters deploy sparsity aware coefficient adaptation, and thereby achieve significant improvement in performance, both in terms of convergence speed and steady-state Excess Mean Square Error (EMSE). A prominent category in this context is the zeroAttracting (ZA) family, in particular, the ZA-LMS \cite{L1LMS} and the ZA-NLMS \cite{L1NLMS} algorithms, where a $\ell_1$-norm penalty of the coefficient vector is added to the LMS/NLMS cost function. However, as the zero-attraction is applied uniformly to all coefficients, if the system is less sparse, there will be zero attraction on the active taps (i.e., taps corresponding to the non-zero coefficients of the system impulse response) also, which will deteriorate the performance. To overcome this problem, a reweighted version of the ZA-LMS/NLMS (RZA-LMS/RZA-NLMS) \cite{L1LMS, L1NLMS} has been proposed which tries to restrict the shrinkage mostly to the inactive (i.e., zero-valued) taps. Motivated from these works, for beamforming applications, a $\ell_1$-norm constrained LMS ($\ell_1$-LMS) algorithm \cite{L1CLMS} is proposed by incorporating $\ell_1$-norm penalty into the CLMS cost function which was later extended to the $\ell_1$-norm Constrained Normalized LMS ($\ell_1$-CNLMS) \cite{L1CNLMS} and $\ell_1$-norm Weighted Constrained Normalized LMS ($\ell_1$-CNLMS) \cite{L1CNLMS}. In the second part of this work, we extend the error non-linear concepts to the sparse case to derive robust sparsity-aware error non-linear adaptive algorithms for adaptive beamforming. Our main contributions include:
\begin{enumerate}
\item We propose CLMLS algorithm by combining LMLS and CLMS, and analyze its performance in detail. 
\item We validate the correctness of the analysis through detailed Monte-Carlo simulations. 
\item We extend the CLMLS to sparse case to derive $\ell_1$-CLMLS and $\ell_1$-WCLMLS algorithms.
\item We demonstrate the superiority of the proposed algorithms over the state-of-the-art by considering system identification and adaptive beam forming applications. 
\end{enumerate}

The rest of the paper is organized as follows. We present the proposed CLMLS algorithm in Section II. Section III deals with the performance analysis of the proposed algorithm. In Section IV, we extend the error non-linear concepts to sparse case to derive $\ell_1$-norm Constrained LMLS algorithm. We present the detailed simulation results in Section V and conclude the paper in Section VI.
\section{Constrained Least Mean Logarithmic Squares (CLMLS)}
In this section, we derive the Constrained Least Mean Logarithmic Squares (CLMLS) algorithm for solving the linearly constrained filtering problems. In the linearly constrained adaptive filtering, the constraints are given by the following set of $K$ equations \cite{Wernerb}:
\begin{equation}\label{1.1}
\centering
{\bf C}^{T} {\bf w} = {\bf z},
\end{equation}
where ${\bf C} \in \mathbb{R}^{L \times K}$ is an $L \times K$ constraint matrix, ${\bf z} \in \mathbb{R}^{K \times 1}$ is a vector containing the $K$ constraint values.

Let the input signal vector ${\bf u}(n) \in \mathbb{R}^{L \times 1}$,  desired signal $d(n) \in \mathbb{R}$ and estimation error of adaptive filter $e(n) \in \mathbb{R}$, then the linear constrained minimization problem in Least Mean Logarithmic Square sense can be stated as
\begin{equation}\label{1.2}
\centering
\min\limits_{\bf w} E\Big[\big|e(n)\big|^{2}- \frac{1}{\alpha} \log\left(1+\alpha \hspace{1mm}e^{2}(n)\right)\Big]  \hspace{0.3cm} s.t. \hspace{0.3cm} {\bf C}^{T} {\bf w} = {\bf z},
\end{equation}
where $\alpha$ is the design parameter \cite{LMLS}, $e(n)= d(n)- {\bf w}^{T} u(n)$. By employing the Lagrange multiplier $\boldsymbol{\lambda}$, the constraints can be included into the objective function, we then have
\begin{equation}\label{1.3}
\centering
J({\bf w})= E\Big[\big|e(n)\big|^{2}- \frac{1}{\alpha} \log\left(1+\alpha \hspace{1mm}e^{2}(n)\right)\Big] - \boldsymbol{\lambda}^{T} \big( {\bf z}- {\bf C}^{T} {\bf w} \big) .
\end{equation}
The solution for $J({\bf w})$ can be obtained in terms of steepest descent iteration as follows:
\begin{equation}\label{1.4}
\centering
{\bf w}(n+1)= {\bf w}(n) - \frac{\mu}{2} \hspace{1mm}\widehat{\bigtriangledown}_{{\bf w}} {\bf J}({\bf w}),
\end{equation}
where $\mu$ is the adaptation step size and the gradient vector $\hat{\bigtriangledown}_{{\bf w}} J({\bf w})$ is given by
\begin{equation}
\hat{\bigtriangledown}_{{\bf w}} J({\bf w}) = - \hspace{0.5mm} \frac{2 \hspace{1mm} \alpha \hspace{0.5mm} e^{3}(n)}{1+ \alpha \hspace{0.5mm} e^{2}(n)} \hspace{1mm} {\bf u}(n) + {\bf C}  \hspace{1mm} \boldsymbol{\lambda}.
\end{equation}
By Pre-multiplying the LHS and RHS of \eqref{1.4} by ${\bf C}^{T}$ and utilizing the constraint relation ${\bf C}^{T} {\bf w}(n+1) = {\bf z}$, the solution for $\boldsymbol{\lambda}$ can be easily obtained. Thus, the update equation of the CLMLS algorithm is given by,
\begin{equation}\label{CLMLS}
\centering
{\bf w}(n+1)= {\bf P} \left({\bf w}(n) + \mu \hspace{1mm} \frac{\alpha \hspace{0.5mm} e^{2}(n)}{1+ \alpha \hspace{0.5mm} e^{3}(n)} \hspace{0.5mm}{\bf u}(n)  \right) + {\bf f},
\end{equation}
where 
\begin{equation}
\begin{split}
{\bf P}&= \left({\bf I}_{L} - {\bf C} \hspace{0.5mm} \Big({\bf C}^{T} \hspace{1mm} {\bf C}\Big)^{-1} \hspace{0.5mm} {\bf C}^{T}\right),\\
{\bf f}&={\bf C} \hspace{0.5mm} \Big({\bf C}^{T}{\bf C}\Big)^{-1} \hspace{1mm}  {\bf z}.
\end{split}
\end{equation}
\section{Performance Analysis}
The performance analysis of the proposed CLMLS algorithm is carried out using the energy conservation approach \cite{Sayedb}. For this, we assume the following:
\begin{itemize}[]
\item[]A1). The input signal ${\bf u}(n)$ is zero-mean Gaussian with covariance matrix ${\bf R}$, is a positive-definite matrix. The observation noise $\vartheta(n)$ is zero-mean i.i.d. Gaussian with variance $\sigma^{2}_{\vartheta}$ and assumed to be independent of input signal ${\bf u}(m)$ for all $n$, $m$.
\end{itemize} 

Under the assumption A1, the optimal filter coefficient vector ${\bf w}_{o}$ is given by \cite{Dinizb},
\begin{equation}\label{2.1}
\centering
{\bf w}_{o}= {\bf h} + {\bf R}^{-1} \hspace{0.5mm} {\bf C} \hspace{0.5mm} \big( {\bf C}^{T} \hspace{0.5mm} {\bf R} \hspace{0.5mm} {\bf C}\big)^{-1} \hspace{0.7mm} \big( {\bf z} -{\bf C}^{T} \hspace{0.5mm}{\bf h}\big),
\end{equation}
where ${\bf h}= {\bf R}^{-1} {\bf p}$. By defining the weight deviation vector as $\widetilde{{\bf w}}(n)= {\bf w}_{o}- {\bf w}(n)$ and recalling the fact that ${\bf w}_{o} - {\bf P}{\bf w}_{o} - {\bf f}={\bf 0}_{L\times 1}$, the recursion of the CLMLS weight deviation vector can then be given as
\begin{equation}\label{2.2}
\centering
\widetilde{{\bf w}}(n+1)= {\bf P} \hspace{0.5mm} \widetilde{{\bf w}}(n) - \mu \hspace{0.5mm} g(e(n)) \hspace{0.5mm} {\bf P} \hspace{0.5mm}{\bf u}(n),
\end{equation}
where $g\big(e(n)\big)=\frac{\alpha \hspace{0.5mm} e^{3}(n)}{1+ \alpha \hspace{0.5mm} e^{2}(n)}$. Since the matrix ${\bf P}$ is idempotent (i.e., ${\bf P}^{2}={\bf P}$), we will have ${\bf P} \hspace{0.5mm}\widetilde{{\bf w}}(n) = \widetilde{{\bf w}}(n)$, thus, one can then obtain
\begin{equation}\label{2.3}
\centering
\widetilde{{\bf w}}(n+1)=  \widetilde{{\bf w}}(n) - \mu \hspace{0.5mm} g\big(e(n)\big) \hspace{0.5mm} {\bf P} \hspace{0.5mm}{\bf u}(n).
\end{equation}
The above recursion serves as the basis for the performance analysis of the CLMLS algorithm.
\subsection{Mean Square Analysis}\label{Analysis}
For any semi positive definite weight matrix ${\boldsymbol{\Sigma}}$, the mean square of the weight deviation vector $\widetilde{{\bf w}}(n)$ satisfies the following enery conservation relation \cite{Sayed2003}:
\begin{equation}\label{2.1.1}
\begin{split}
E\big[\|\widetilde{{\bf w}}(n+1)\|_{\boldsymbol{\Sigma}}^{2}\big]&=  E \big[\|\widetilde{{\bf w}}(n)\|_{\boldsymbol{\Sigma}}^{2}\big] - 2 \hspace{0.5mm} \mu \hspace{0.5mm} E \big[ e^{{\bf P} \hspace{0.5mm} \boldsymbol{\Sigma}}_{a}(n) \hspace{0.5mm} g(e(n)) \big] \\[2mm]
& \hspace{5mm} + \mu^{2}  \hspace{0.5mm} E\big[g^{2}(e(n)) \hspace{0.5mm} \|{\bf u}(n)\|^{2}_{{\bf P} \hspace{0.5mm} \boldsymbol{\Sigma} \hspace{0.5mm} {\bf P}} \big],
\end{split}
\end{equation}
where $e^{{\bf P} \hspace{0.5mm} \boldsymbol{\Sigma}}_{a}(n)=\widetilde{{\bf w}}^{H}(n) \hspace{0.5mm} \boldsymbol{\Sigma} \hspace{0.7mm} {\bf P} \hspace{0.5mm} {\bf u}(n)$ is the weighted \emph{a priori} estimation error. To simplify the above, following the same lines of \cite{LMLS}, at this stage we assume the following (which are commonly used in the analysis of
adaptive filters with error non-linearities \cite{Sayed2003}):
\begin{itemize}
\item[]A2). The \emph{a priori} estimation error $e_{a}(n)$ has Gaussian distribution and it is jointly Gaussian with the weighted \emph{a priori} estimation error $e^{{\bf P} \hspace{0.5mm} \boldsymbol{\Sigma}}_{a}(n)$ for any constant matrices $\boldsymbol{\Sigma}$ and ${\bf P}$. This assumption is reasonable for long filters, i.e., for large $L$ and  sufficiently small step size value $\mu$. \\[-2mm]
\item[]A3). The random variables $\|{\bf u}(n)\|^{2}_{{\bf P} \hspace{0.5mm} \boldsymbol{\Sigma} \hspace{0.5mm} {\bf P}}$ and $g^{2}(e(n))$ are uncorrelated, which results 
\begin{equation}\label{2.1.2}
\hspace{-4mm}E\big[g^{2}(e(n)) \hspace{0.5mm} \|{\bf u}(n)\|^{2}_{{\bf P} \hspace{0.5mm} \boldsymbol{\Sigma} \hspace{0.5mm} {\bf P}} \big]= E\big[g^{2}(e(n))\big] \hspace{0.5mm} E\big[\|{\bf u}(n)\|^{2}_{{\bf P} \hspace{0.5mm} \boldsymbol{\Sigma} \hspace{0.5mm} {\bf P}} \big]
\end{equation}
\end{itemize} 
\subsection{Transient Performance}
Under the the assumption A$1$, and A$2$, the estimation error $e(n)= e_{a}(n) + \vartheta(n)$ is Gaussian distributed (which is reasonable, as it is generated from the summation of two independent Gaussian distributed random variables). Hence, same as Lemma$1$ in \cite{LMLS}, under the assumptions A1, A2 and using the Prices's Theorem \cite{Price}, we can write,
\begin{equation}\label{2.2.1}
E \big[ e^{{\bf P} \hspace{0.5mm} \boldsymbol{\Sigma}}_{a}(n)\hspace{1mm}  g(e(n))\big]= E \big[ e^{{\bf P} \hspace{0.5mm} \boldsymbol{\Sigma}}_{a}(n)\hspace{1mm}  e_{a}(n)\big] \hspace{1mm} \frac{E[e(n) \hspace{1mm} g(e(n))]} {E[e^{2}(n)]}
\end{equation}
Substituting the \eqref{2.1.2} and \eqref{2.2.1} in \eqref{2.1.1}, we obtain
\begin{equation}\label{2.2.2}
\begin{split}
E\big[\|\widetilde{{\bf w}}(n+1)\|_{\boldsymbol{\Sigma}}^{2}\big]&= 
 E \big[\|\widetilde{{\bf w}}(n)\|_{\boldsymbol{\Sigma}}^{2}\big] \\
&\hspace{5mm}- 2 \hspace{0.5mm} \mu \hspace{0.5mm} E \big[ e^{{\bf P} \hspace{0.5mm} \boldsymbol{\Sigma}}_{a}(n)\hspace{1mm}  e_{a}(n)\big] \hspace{1mm} h_{G}(n) \\
&\hspace{5mm}+ \mu^{2}  \hspace{0.5mm} E\big[\|{\bf u}(n)\|^{2}_{{\bf P} \hspace{0.5mm} \boldsymbol{\Sigma} \hspace{0.5mm} {\bf P}} \big] \hspace{1mm} h_{U}(n),
\end{split}
\end{equation}
where 
\begin{equation}
\begin{split}
h_{G}(n)&=\frac{E[e(n) \hspace{1mm} g(e(n))]} {E[e^{2}(n)]},\\[2mm]
h_{U}(n)&=E\big[g^{2}(e(n))\big].
\end{split}
\end{equation}
These functions are similar to the  ones presented in \cite{LMLS} and can be evaluated using the same procedure. Since $E\big[ e^{{\bf P} \hspace{0.5mm} \boldsymbol{\Sigma}}_{a}(n)\hspace{1mm}  e_{a}(n)\big]= E\big[ \widetilde{{\bf w}}^{T}(n) \hspace{0.5mm} \boldsymbol{\Sigma} \hspace{0.5mm} {\bf P} \hspace{0.5mm} {\bf u}^{T}(n) \hspace{0.5mm} {\bf u}(n) \hspace{0.5mm} \widetilde{{\bf w}}(n) \big]= E\big[ \widetilde{{\bf w}}^{T}(n) \hspace{0.5mm} \boldsymbol{\Sigma} \hspace{0.5mm} {\bf P} \hspace{0.5mm} E\big[{\bf u}^{T}(n) \hspace{0.5mm} {\bf u}(n) \big]\hspace{0.5mm} {\bf P} \hspace{0.5mm} \widetilde{{\bf w}}(n) \big]= E\big[ \widetilde{{\bf w}}^{T}(n) \hspace{0.5mm} \boldsymbol{\Sigma} \hspace{0.5mm} {\bf P} \hspace{0.5mm} {\bf R} \hspace{0.5mm} $ \newline $ {\bf P} \hspace{0.5mm} \widetilde{{\bf w}}(n) \big] = E\big[\|\widetilde{{\bf w}}(n)\|^{2}_{\boldsymbol{\Sigma} \hspace{0.5mm} {\bf P} \hspace{0.5mm} {\bf R} \hspace{0.5mm} {\bf P}} \big]$  and $E\big[\|{\bf u}(n)\|^{2}_{{\bf P} \hspace{0.5mm} \boldsymbol{\Sigma} \hspace{0.5mm} {\bf P}} \big]=E\big[{\bf u}^{T}(n) \hspace{0.5mm} {\bf P} \hspace{0.5mm} \boldsymbol{\Sigma} \hspace{0.5mm} {\bf P} \hspace{0.5mm} {\bf u}(n) \big]=Trace\big( {\bf P} \hspace{0.5mm} {\bf R} \hspace{0.5mm} {\bf P} \hspace{0.5mm} \boldsymbol{\Sigma}\big)$, the above recursion becomes
\begin{equation}\label{2.2.3}
\begin{split}
&E\big[\|\widetilde{{\bf w}}(n+1)\|_{\boldsymbol{\Sigma}}^{2}\big]=\\[1mm]
&\hspace{7mm}  E \big[\|\widetilde{{\bf w}}(n)\|_{\boldsymbol{\Sigma}^{'}(n)}^{2}\big] + \mu^{2}  \hspace{0.5mm} Trace\big({\bf P} \hspace{0.5mm} {\bf R} \hspace{0.5mm} {\bf P} \hspace{0.5mm} \boldsymbol{\Sigma} \big) \hspace{1mm} h_{U}(n),
\end{split}
\end{equation}
where 
\begin{equation}\label{2.2.4}
{\boldsymbol{\Sigma}^{'}}(n)= \boldsymbol{\Sigma}- 2 \hspace{0.5mm} \mu \hspace{0.5mm} h_{G}(n) \hspace{0.5mm} \boldsymbol{\Sigma} \hspace{1mm} {\bf P} \hspace{0.5mm} {\bf R} \hspace{0.5mm} {\bf P}.
\end{equation}
To extract the matrix $\boldsymbol{\Sigma}$ from the expectation
terms, a weighted variance relation is introduced by using $L^{2}
\times 1$ column vectors $\boldsymbol{\sigma}=
\textit{vec}\{\boldsymbol{\Sigma}\}$ and  $\boldsymbol{\sigma^{'}}(n)=
\textit{vec}\{\boldsymbol{\Sigma}^{'} (n)\}$, where
$\textit{vec}\{\cdot\}$ denotes the vector operator. In addition,
$\textit{vec}\{\cdot\}$ is also used to recover the original matrix
$\boldsymbol{\Sigma}$ from $\boldsymbol{\sigma}$. One property of
the $\textit{vec}\{\cdot\}$ operator when working with the Kronecker
product \cite{BlockKronecker} is used in this work, namely,
\begin{equation}\label{2.2.5}
\begin{split}
\textit{vec}\{{\bf A} \boldsymbol{\Sigma} {\bf B}\}= ( {\bf B}^{T}
\otimes {\bf A} ) \hspace{1.2mm} \boldsymbol{\sigma},
\end{split}
\end{equation}
where ${\bf A} \otimes {\bf B}$ indicates the Kronecker product of two matrices. Using the above, after vectorization of \eqref{2.2.4}, a linear relation between the corresponding vectors $\{\boldsymbol{\sigma},\boldsymbol{\sigma}^{'}(n)\}$ can be formulated as follows:
\begin{equation}\label{2.2.6}
\begin{split}
\boldsymbol{\sigma}^{'}(n)= \textbf{F}(n) \hspace{1mm} \boldsymbol{\sigma},
\end{split}
\end{equation}
where $\textbf{F}(n)$ is a $L^{2}\times L^{2}$ matrix and defined as,
\begin{equation}\label{2.2.7}
\begin{split}
\textbf{F}(n)={\bf I}_{L^{2}} - 2 \mu \hspace{1mm} h_{G}(n) \hspace{1mm} \big({\bf I}_{L} \otimes {\bf P} \big) \hspace{1mm} \big({\bf I}_{L} \otimes {\bf R} \big) \hspace{1mm} \big({\bf I}_{L} \otimes {\bf P} \big). 
\end{split}
\end{equation}

The second term in the RHS of \eqref{2.2.4} can be simplified as
\begin{equation}\label{2.2.8}
\mu^{2}  \hspace{0.5mm} Trace\big({\bf P} \hspace{0.5mm} {\bf R} \hspace{0.5mm} {\bf P} \hspace{0.5mm} \boldsymbol{\Sigma} \big) \hspace{1mm} h_{U}(n) = \mu^{2} \hspace{0.5mm} h_{U}(n) \hspace{1mm} \boldsymbol{\gamma}^{T} \hspace{0.5mm} \boldsymbol{\sigma},
\end{equation}
where $\boldsymbol{\gamma}=\textit{vec}\big\{{\bf P} \hspace{0.5mm} {\bf R} \hspace{0.5mm} {\bf P}\big\} = \big({\bf P} \otimes {\bf P}\big) \hspace{1mm} \boldsymbol{\gamma}_{{\bf R}}$, with $\boldsymbol{\gamma}_{{\bf R}}= \textit{vec}\{{\bf R}\}$.

Substituting \eqref{2.2.6} and \eqref{2.2.8} in \eqref{2.2.2}, we obtain
\begin{equation}\label{2.2.9}
\begin{split}
E\big[\|\widetilde{{\bf w}}(n+1)\|_{\boldsymbol{\sigma}}^{2}\big]=  E \big[\|\widetilde{{\bf w}}(n)\|_{{\bf F}(n) \hspace{0.5mm}\boldsymbol{\sigma}}^{2}\big] + \mu^{2} \hspace{0.5mm} h_{U}(n) \hspace{1mm}\boldsymbol{\gamma}^{T} \hspace{0.5mm} \boldsymbol{\sigma},
\end{split}
\end{equation}
Iterating the recursion \eqref{2.2.9}, starting from $n = 0$, we obtain
\begin{equation}\label{2.2.10}
\begin{split}
E\big[\|\widetilde{{\bf w}}(n+1)\|_{\boldsymbol{\sigma}}^{2}\big]&=  E \Big[\|\widetilde{{\bf w}}(0)\|_{\left(\prod\limits_{i=0}^{n}{\bf F}(i) \right)\hspace{0.5mm} \boldsymbol{\sigma}}^{2} \Big] \\
&+ \mu^{2} \hspace{0.2mm}\boldsymbol{\gamma}^{T} \left(\sum\limits_{i=0}^{n} h_{U}(n)  \hspace{0.2mm} \left(\prod\limits_{j=i+1}^{n}{\bf F}(j) \right)\right)\hspace{0.2mm} \boldsymbol{\sigma}.
\end{split}
\end{equation}
By relating $E\|\widetilde{{\bf w}}(n+1)\|_{\boldsymbol{\sigma}}^{2}$ and $E\|\widetilde{{\bf w}}(n)\|_{\boldsymbol{\sigma}}^{2}$, we can then have
\begin{equation}\label{2.2.11}
\begin{split}
&E\|\widetilde{{\bf w}}(n+1)\|_{\boldsymbol{\sigma}}^{2}= \\
&\hspace{5mm} E\|\widetilde{{\bf w}}(n)\|_{\boldsymbol{\sigma}}^{2} - \|\widetilde{{\bf w}}(0)\|^{2}_{\left({\bf I}_{L^{2}}-\textbf{F}(n)\right)\hspace{0.7mm} \left(\prod\limits_{i=0}^{n}{\bf F}(i) \right)\hspace{0.5mm} \boldsymbol{\sigma}} \\
&\hspace{5mm}+ \mu^{2} \hspace{0.5mm} h_{U}(n) \hspace{0.7mm} \boldsymbol{\gamma}^{T} \hspace{0.5mm} \boldsymbol{\sigma} \\
& \hspace{5mm}+ \mu^{2} \hspace{0.5mm} \boldsymbol{\gamma}^{T} \left(\sum\limits_{i=0}^{n-1} h_{U}(i) \hspace{0.3mm} \big( \textbf{F}(n)-{\bf I}_{L^{2}}\big)\left(\prod\limits_{j=i+1}^{n-1}{\bf F}(j) \right) \right) \boldsymbol{\sigma}.
\end{split}
\end{equation}
This weighted variance relation is helpful to characterize the transient behavior of the proposed CLMLS algorithm. By evaluating $\sigma^{2}_{e(n)}$ at each index $n$ through $\sigma^{2}_{e(n)}=\sigma^{2}_{e_{a}(n)}+\sigma^{2}_{\vartheta}$, the functions $h_{G}(n)$ and $h_{U}(n)$, which are functions of $\sigma^{2}_{e(n)}$ can be evaluated. By choosing $\boldsymbol{\Sigma}={\bf R}$, the transient EMSE (i.e., $\zeta(n) = E\|\widetilde{{\bf w}}(n)\|_{{\bf R}}^{2}$) performance curves of the proposed CLMLS algorithm can then be obtained as
\begin{equation}\label{2.2.12}
\begin{split}
&\zeta(n)= \\
&\zeta(n) - \|\widetilde{{\bf w}}(0)\|^{2}_{\left({\bf I}_{L^{2}}-\textbf{F}(n)\right)\hspace{0.7mm} \left(\prod\limits_{i=0}^{n}{\bf F}(i) \right)\hspace{0.5mm} vec\{{\bf R}\}} \\
&+ \mu^{2} \hspace{0.5mm} h_{U}(n) \hspace{0.7mm} \boldsymbol{\gamma}^{T} \hspace{0.5mm} vec\{{\bf R}\} \\
&+ \mu^{2} \hspace{0.5mm} \boldsymbol{\gamma}^{T} \left(\sum\limits_{i=0}^{n-1} h_{U}(i) \hspace{0.3mm} \big( \textbf{F}(n)-{\bf I}_{L^{2}}\big)\left(\prod\limits_{j=i+1}^{n-1}{\bf F}(j) \right)  \right) vec\{{\bf R}\}.
\end{split}
\end{equation} 
Note that by choosing $\boldsymbol{\Sigma}={\bf I}_{L}$, the transient MSD (i.e., $\xi(n)= E\|\widetilde{{\bf w}}(n)\|^{2}$)  performance curves can be obtained.
\subsection{Steady-state Performance}
For large $n$, i.e., in steady-state, we will have $\lim\limits_{n \to \infty} E\big[\|\widetilde{{\bf w}}(n+1)\|_{\boldsymbol{\Sigma}}^{2}\big]=\lim\limits_{n \to \infty} E\big[\|\widetilde{{\bf w}}(n)\|_{\boldsymbol{\Sigma}}^{2}\big]$. Then, from \eqref{2.2.9}, we can then have
\begin{equation}\label{2.3.1}
\begin{split}
\lim\limits_{n \to \infty} E\big[\|\widetilde{{\bf w}}(n)\|_{\big({\bf I}_{L^{2}} -{\bf F}(n) \big) \boldsymbol{\sigma}}^{2}\big]&=  \mu^{2} \hspace{0.5mm}\lim\limits_{n \to \infty} h_{U}(n) \hspace{1mm}\boldsymbol{\gamma}^{T} \hspace{0.5mm}  \boldsymbol{\sigma}.
\end{split}
\end{equation}
By choosing $\boldsymbol{\sigma}=\big({\bf I}_{L^{2}} -{\bf F}(n) \big)^{-1} \hspace{0.5mm}vec\{{\bf R}\}$, we obtain the steady-state EMSE of CLMLS, i.e., $\zeta(\infty)=\lim\limits_{n \to \infty} E[e^{2}_{a}(n)]$, which is given by,
\begin{equation}\label{2.3.2}
\begin{split}
\zeta(\infty)= \mu^{2} \hspace{0.5mm} \lim\limits_{n \to \infty} h_{U}(n) \hspace{1mm} \boldsymbol{\gamma}^{T} \hspace{0.5mm} \big({\bf I}_{L^{2}} -{\bf F}(n) \big)^{-1} \hspace{0.5mm} vec\{{\bf R}\}.
\end{split}
\end{equation}
After some simplifications, we obtain,
\begin{equation}\label{2.3.3}
\begin{split}
\zeta(\infty)= \frac{\mu}{2} \hspace{0.5mm} \lim\limits_{n \to \infty} \frac{h_{U}(n)}{h_{G}(n)} \hspace{1mm} \boldsymbol{\gamma}^{T} \hspace{0.5mm} {\bf S}^{-1} \hspace{0.5mm} vec\{{\bf R}\},
\end{split}
\end{equation}
where ${\bf S}=\big({\bf P}  \otimes  {\bf I}_{L}\big) \hspace{1mm} \big({\bf R} \otimes  {\bf I}_{L} \big) \hspace{1mm} \big({\bf P}  \otimes  {\bf I}_{L} \big)$.
\begin{itemize}
\item[]A$4$). For an appropriate value of $\mu$, in steady-state, simiar to \cite{LMLS}, we assume
\begin{subequations}\label{2.3.4a}
\begin{align}\label{2.3.4a}
\begin{split}
h_{G}&= \lim\limits_{n \to \infty} \hspace{1mm} \frac{1}{E[e^{2}(n)]}  \hspace{1mm}  E\left[ \frac{\alpha \hspace{0.5mm} e^{4}(n)} {1+\alpha \hspace{0.5mm} e^{2}(n)} \right] \hspace{2cm}\\
&= \frac{\alpha}{\sigma^{2}_{e}} \lim\limits_{n \to \infty} \hspace{1mm} E[e^{4}(n)],
\end{split}
\end{align}
and
\begin{align}\label{2.3.4b}
\begin{split}
h_{U}&= \lim\limits_{n \to \infty} \hspace{1mm} E\left[ \frac{\alpha^{2} \hspace{0.5mm} e^{6}(n)} {\big(1+\alpha \hspace{0.5mm} e^{2}(n) \hspace{0.5mm} \big)^{2}} \right] = \alpha^{2} \lim\limits_{n \to \infty} \hspace{1mm} E[e^{6}(n)],
\end{split}
\end{align}
\end{subequations}
where $\sigma^{2}_{e}=\lim\limits_{n \to \infty}E[e^{2}(n)]$. 
\end{itemize}

Using A$4$, from \eqref{2.3.3}, the steady-state EMSE is given by
\begin{equation}\label{2.3.5}
\begin{split}
\zeta(\infty) &= \frac{\mu}{2} \hspace{1mm}  \alpha \hspace{1mm}  \sigma^{2}_{e} \hspace{1mm} \lim\limits_{n \to \infty} \hspace{1mm} \frac{E[e^{6}(n)]}{E[e^{4}(n)]} \hspace{1mm} \boldsymbol{\gamma}^{T} \hspace{0.5mm} {\bf S}^{-1} \hspace{0.5mm} vec\{{\bf R}\}\\
&= \frac{\mu}{2} \hspace{1mm}  \alpha \hspace{1mm} \sigma^{2}_{e} \hspace{1mm} \frac{15 \hspace{1mm} \sigma^{6}_{e}}{3 \hspace{1mm} \sigma^{4}_{e}} \hspace{1mm} \boldsymbol{\gamma}^{T} \hspace{0.5mm} {\bf S}^{-1} \hspace{0.5mm} vec\{{\bf R}\}.
\end{split}
\end{equation}
By substituting $\sigma^{2}_{e}= \zeta(\infty) +\sigma^{2}_{\vartheta}$, we can then have
\begin{equation}\label{2.3.6}
\begin{split}
\zeta(\infty) = \frac{5 \hspace{1mm}\mu}{2} \hspace{1mm}  \alpha \hspace{1mm}  \big( \zeta(\infty)+  \sigma^{2}_{\vartheta}\big)^{2}   \hspace{1mm}  \beta,
\end{split}
\end{equation}
where $\beta=  \boldsymbol{\gamma}^{T} \hspace{0.5mm} {\bf S}^{-1} \hspace{0.5mm} vec\{{\bf R}\}$. After some simple algebra, steady-state EMSE of CLMLS algorithm is,
\begin{equation}\label{2.3.7}
\begin{split}
\zeta(\infty) = \frac{1- 5 \hspace{0.7mm}\alpha \hspace{0.7mm} \mu \hspace{0.7mm} \beta \hspace{0.7mm} \sigma^{2}_{\vartheta} \pm \sqrt{1-10 \hspace{0.7mm} \alpha \hspace{0.7mm} \mu \hspace{0.7mm} \beta \hspace{0.7mm} \sigma^{2}_{\vartheta}}}{5 \hspace{0.7mm} \alpha \hspace{0.7mm} \mu \hspace{0.7mm} \beta }.
\end{split}
\end{equation}
Note that by choosing $\boldsymbol{\sigma}=\big({\bf I}_{L^{2}} -{\bf F}(n) \big)^{-1}$, we obtain the steady-state MSD of CLMLS, i.e., $\xi(\infty)=\lim\limits_{n \to \infty} E\big[\|\widetilde{{\bf w}}(n)\|^{2}\big]$.

%\begin{remark}
%Under steady state , a sufficient condition for the convergence of the above Mean Square Deviation (MSD) recursion is,
%\begin{equation}
%\lambda_{max}\Big( {\bf I}_{L}- 2 \hspace{0.5mm} \mu \hspace{0.5mm} {\bf P} \hspace{0.5mm} {\bf R} \hspace{0.5mm} {\bf P}\hspace{1mm} h_{G}(n) \Big), 
%\end{equation}
%where $\lambda_{max}(.)$ denotes the maximum eigenvalue of its argument. After some simple manipulations, it is easy to see that the above condition gets satisfied for $0 < \mu < \frac{1}{\lambda_{max}\Big({\bf P} \hspace{0.5mm} {\bf R} \hspace{0.5mm} {\bf P}\Big) \hspace{0.5mm} h_{G}(n)}$.
%\end{remark}

\section{$\ell_1$-norm Linearly Constrained LMLS Algorithm}
Inspired by the LASSO \cite{LASSO} and the sparse LMS algorithms \cite{L1LMS}, a $\ell_{1}$-norm constraint based CLMS algorihtm is proposed in \cite{L1CLMS}. The $\ell_{1}$-CLMS algorithm incorporates the $\ell_{1}$ penalty into the cost function of CLMS thereby achieves improved performance over the CLMS for identifying the sparse system. In order to exploit the underlying system sparsity, the $\ell_{1}$-norm penalty can also be added to the list of constraints in \eqref{1.2} and the corresponding cost function is given by
\begin{equation}\label{5.1}
\begin{split}
J({\bf w})&= E\Big[\big|e(n)\big|^{2}- \frac{1}{\alpha} \log\left(1+\alpha \hspace{1mm}e^{2}(n)\right)\Big] - \boldsymbol{\lambda}_{1}^{T} \big( {\bf z}- {\bf C}^{T} {\bf w} \big) \\[2mm]
& \hspace{5mm} - \lambda_{2} \big( t - \|{\bf w}\|_{1} \big).
\end{split}
\end{equation}  
Using the steepest descent method, at each iteration, the coefficient vector is then updated as
\begin{equation}\label{5.2}
\centering
{\bf w}(n+1)= {\bf w}(n) - \frac{\mu}{2} \hspace{1mm}\widehat{\bigtriangledown}_{{\bf w}} {\bf J}({\bf w}),
\end{equation}
where $\hat{\bigtriangledown}_{{\bf w}} J({\bf w}) = -2 \hspace{1mm}  g\big(e(n)\big)\hspace{1mm} {\bf u}(n) + {\bf C} \boldsymbol{\lambda}_{1} + \lambda_2 \hspace{1mm} sign\big( {\bf w}\big)$, with $sign(\cdot)$ denoting the basic signum function. Pre-multiplying the LHS and RHS of \eqref{5.2} by ${\bf C}^{T}$ and using the constraint relation ${\bf C}^{T} \hspace{1mm} {\bf w}(n+1) = {\bf C}^{T} \hspace{1mm} {\bf w}(n)= {\bf z}$, the solution for $\boldsymbol{\lambda}_{1}$ can be obtained as
\begin{equation}\label{5.3}
\centering
\boldsymbol{\lambda}_{1}= ({\bf C}^{T} {\bf C})^{-1} \hspace{1mm} {\bf C}^{T} \Big( 2 \hspace{1mm} g\big(e(n)\big) \hspace{1mm} {\bf u}(n) - \lambda_{2} \hspace{1mm} {\bf s}(n) \Big),
\end{equation}
where ${\bf s}(n)=sign \big({\bf w}(n) \big)$. Defining the $\ell_{1}$-norm of the weight vector as $t(n)= {\bf s}^{T}(n) \hspace{1mm} sign \big({\bf w}(n) \big)$, Pre-multiplying the LHS and RHS of \eqref{5.3} by $s^{T}$ and using the constraint relation $\| {\bf w} \|_{1}=t$, we will have
\begin{equation}\label{5.4}
\begin{split}
t= t(n)- \frac{\mu}{2}\hspace{0.2mm} \left( \begin{array}{l} - g\big(e(n)\big) \hspace{0.2mm} {\bf s}^{T}(n) \hspace{0.2mm} {\bf u}(n) + {\bf s}^{T}(n) \hspace{0.2mm} {\bf C} \hspace{0.2mm} \boldsymbol{\lambda}_{1} \\[2mm]
+ \lambda_{2} \hspace{0.2mm} {\bf s}^{T}(n) \hspace{0.2mm} {\bf s}(n) \end{array}\right).
\end{split}
\end{equation}
By denoting $e_{L_{1}}(n)= t - t(n)$ and rearranging the terms, $\lambda_{2}$ can be obtained as
\begin{equation}\label{5.5}
\centering
\lambda_{2}=  \frac{1}{N} \Big( - \hspace{0.1mm} \frac{2}{\mu} \hspace{0.2mm} e_{L_{1}}(n) +   2 \hspace{0.2mm} g\big(e(n)\big) \hspace{0.2mm} {\bf s}^{H}(n) \hspace{0.2mm} {\bf u}(n) -    {\bf s}^{H}(n) \hspace{0.2mm} {\bf C} \hspace{0.2mm} \boldsymbol{\lambda}_{1} \Big),
\end{equation}
where $N= {\bf s}^{T}(n) {\bf s}(n)$. After solving the \eqref{5.4} and \eqref{5.3} to obtain the Lagrangian multipliers $\boldsymbol{\lambda}_{1}$ and $\lambda_{2}$, the weight update equation of $\ell_{1}$-CLMLS algorithm can then be obtained as
\begin{equation}\label{CLMLS}
\centering
{\bf w}(n+1)= {\bf P} \left({\bf w}(n) + \mu \hspace{1mm}  g\big(e(n)\big) \hspace{1mm} {\bf P}^{'}(n) \hspace{1mm} {\bf u}(n)  \right) + {\bf f} + {\bf f}_{L_1}(n),
\end{equation}
where 
\begin{equation}
\begin{split}
{\bf P}^{'}(n)&= \left({\bf I}_{L} - \left(\frac{{\bf P} \hspace{1mm} {\bf s}(n)} {\| {\bf P} \hspace{1mm} {\bf s}(n)\|^{2}_{2}} \right) \hspace{1mm} {\bf s}^{T}(n) \hspace{1mm} \right) {\bf P},\\[2mm]
e_{L_1}(n)&= t- {\bf s}^{T}(n) \hspace{1mm} {\bf w}(n),\\[2mm]
{\bf f}_{L_1}(n)&= e_{L_1}(n) \hspace{1mm} \left(\frac{{\bf P} \hspace{1mm} {\bf s}(n)} {\| {\bf P} \hspace{1mm} {\bf s}(n)\|^{2}_{2}} \right).
\end{split}
\end{equation}
However, as the $\ell_1$-norm penalty uniformly shrinks all the coefficients, if the system is less sparse, the shrinkage on the active taps (i.e., taps corresponding to the non-zero coefficients of the system impulse response) will enhance the misadjustment. To overcome this problem, similar to \cite{L1CLMS}, we also use the reweighted version of the   $\ell_{1}$-norm penalty as the constraint. The objective function then becomes
\begin{equation}\label{5.1}
\begin{split}
J({\bf w})&= E\Big[\big|e(n)\big|^{2}- \frac{1}{\alpha} \log\left(1+\alpha \hspace{1mm}e^{2}(n)\right)\Big] - \boldsymbol{\lambda}_{1}^{T} \big( {\bf z}- {\bf C}^{T} {\bf w} \big) \\[2mm]
& \hspace{5mm} - \lambda_{2} \Big( t - \frac{2}{\pi} \sum\limits_{j=1}^{L} \arctan(\beta |w_{j}|) \Big),
\end{split}
\end{equation} 
where $\beta$ is the slope factor of weight $\ell_{1}$-norm penalty.
Following the same procedure as above, we can obtain the weight update equation of  $\ell_{1}$-WCLMLS as follows:
\begin{equation}\label{CLMLS}
\centering
{\bf w}(n+1)= {\bf P} \left({\bf w}(n) + \mu \hspace{1mm}  g\big(e(n)\big) \hspace{1mm} {\bf P}^{'}(n) \hspace{1mm} {\bf u}(n)  \right) + {\bf f} + {\bf f}_{L_1}(n),
\end{equation}
where 
\begin{equation}
\begin{split}
{\bf P}^{'}(n)&= \left({\bf I}_{L} - \left(\frac{{\bf P} \hspace{1mm} {\bf s}(n)} {\| {\bf P} \hspace{1mm} {\bf s}(n)\|^{2}_{2}} \right) \hspace{1mm} {\bf s}^{T}(n) \hspace{1mm} \right) {\bf P},\\[2mm]
e_{L_1}(n)&= t- {\bf s}^{T}(n) \hspace{1mm} {\bf w}(n),\\[2mm]
{\bf f}_{L_1}(n)&= e_{L_1}(n) \hspace{1mm} \left(\frac{{\bf P} \hspace{1mm} {\bf s}(n)} {\| {\bf P} \hspace{1mm} {\bf s}(n)\|^{2}_{2}} \right),
\end{split}
\end{equation}
with 
\begin{equation}
{\bf s}(n)=\frac{2 \beta}{\pi} \left[\frac{sign(w_{1}(n))}{\beta^{2}|w_{1}(n)|^{2} +1}, \cdots, \frac{sign(w_{L}(n))}{\beta^{2}|w_{L}(n)|^{2} +1} \right]^{T}.
\end{equation}
\section{Simulation Results}
This Section presents the detailed simulation results with two fold objective:
\begin{enumerate}
\item To evaluate and compare the performance of the proposed algorithms with the state-of-the-art 
\item To validate the theoretical results obtained in analysis through Monte-Carlo simulations. 
\end{enumerate}
A series of experiments is conducted for this via system identification and adaptive beam forming applications which are described below:
\paragraph{Experiment $1$}
First we considered a constrained system identification problem, where the filter coefficients are constrained to preserve the linear phase at each iteration. As in \cite{CAPA}, a system of length $L=10$ is considered and to satisfy the linear phase condition, we set 
\begin{equation}
{\bf C}=\left[\begin{array}{l} \hspace{3mm}{\bf  I}_{L/2}\\[3mm]
\hspace{3mm} {\bf 0}^{T}\\[2mm]
-{\bf J}_{L/2} \end{array}\right],
\end{equation}
where ${\bf J}$ being the reversal matrix (an identity matrix with all lines in reversed order) and ${\bf f}=[ 0, \cdots, 0]^{T}$. Input signal is zero mean white Gaussian with unity variance and the observation noise is taken to be zero-mean white Gaussian with  variance $\sigma^{2}_{\vartheta}=0.01$ (i.e., SNR=20 dB). The adaptation step size of LMLS and proposed CLMLS is fixed at $\mu=0.05$ while the $\mu$ of the LMS and CLMS is adjusted such that the steady state MSD of these algorithms is same as that of LMLS and CLMLS, respectively. 

\begin{figure}[h!]
\centering
\includegraphics [height=90mm,width=88mm]{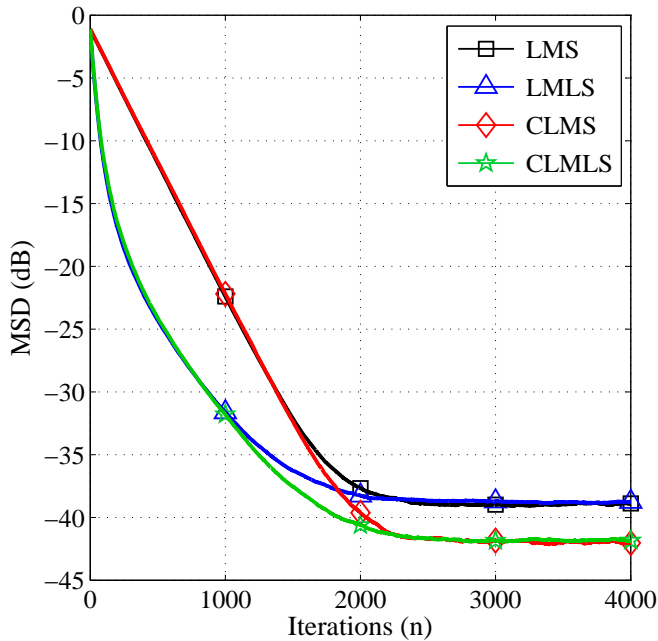}
\caption{Learning curves of the CLMS and the proposed CLMLS algorithms for white Gaussian input with SNR=$20$dB. }
\label{Learning_curves}
\end{figure}

The performance is evaluated by the MSD [ref] defined as $10 \log_{10}
\left(\mathit{E}\left(\frac{\|{\bf w}_{opt}-{\bf
w}(n)\|_{2}^{2}}{\|{\bf w}_{opt}\|^{2}_{2}}\right)\right) $. Ensemble average of $500$ independent trails is used for calculating the MSD. 
The learning curves (i.e., MSD in dB vs no. of iterations) of the proposed CLMLS along with other algorithms are shown in Fig.~\ref{Learning_curves}. It can be observed that the proposed CLMLS clearly outperforms the CLMS algorithm. 

\paragraph{Experiment $2$}
Next, we validate the analytical results presented in the Section~\ref{Analysis}. For this, the proposed CLMLS is simulated to identify the same unknown system used above for different values of SNR \{30 dB, 25 dB, 20 dB\}, i.e, \{{$\sigma_{\vartheta}^{2}=0.001$, $\sigma_{\vartheta}^{2}=0.0031$, $\sigma_{\vartheta}^{2}=0.01$\}. The other parameters remaining same as the above.  The MSD of the proposed CLMLS algorithm  $E[\widetilde{{\bf w}}^{2}(n)]$ is evaluated by averaging $\widetilde{{\bf w}}^{2}(n)$ over $500$ independent trails and plotted in Fig.~\ref{transient_NV}. Similarly, for different adaptation step size values $\{0.03, 0.05, 0.1\}$, the proposed CLMS is simulated and its corresponding MSD is plotted in Fig.~\ref{transient_Muw}. We also evaluated the theoretical MSD using \eqref{2.2.10} and plotted in Fig.~\ref{transient_NV} and Fig.~\ref{transient_Muw}, respectively. From these figures, it can be observed that the theoretical results show good agreement with the simulation results which in turn validates the correctness of the presented analysis.

\begin{figure}[h!]
\centering
 \subfigure[for different values of $\sigma_{\vartheta}^{2}$ \label{transient_NV}]{
\includegraphics[height=85mm,width=88mm]{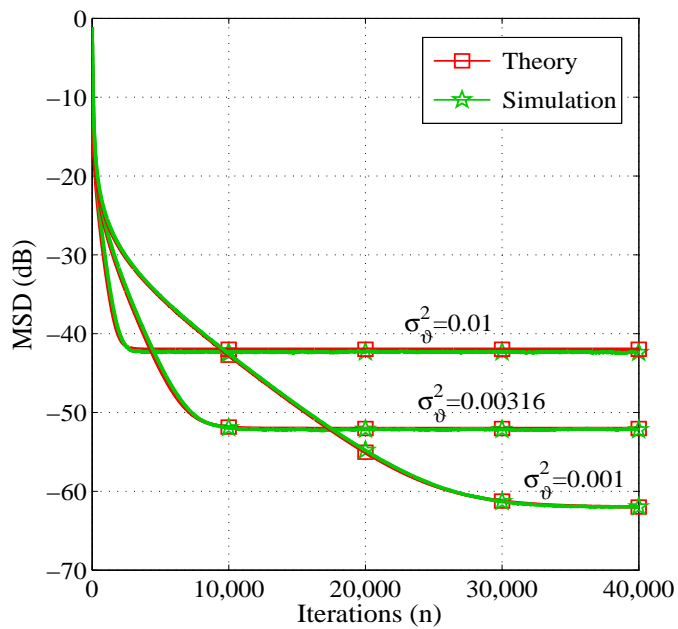} } \\
\subfigure[for different values of $\mu$ \label{transient_Muw}]{
\includegraphics[height=85mm,width=88mm]{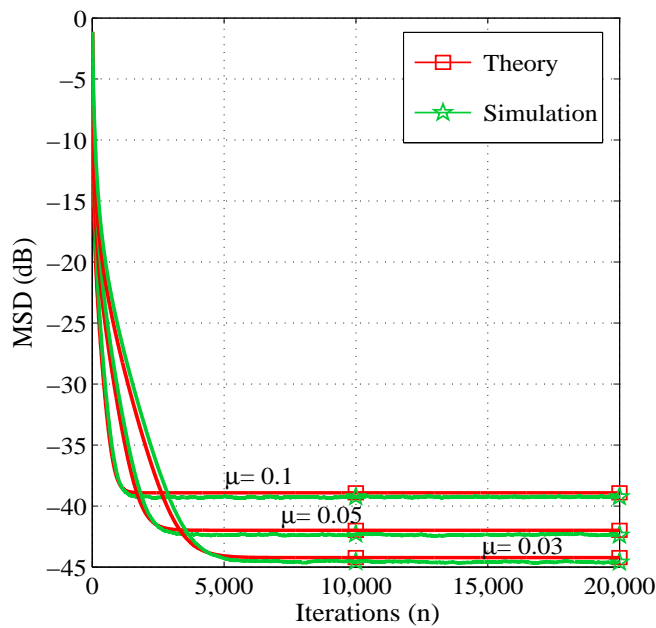} }
\caption{Learning curves of the proposed LMLS for different values of a). SNR, b). Adaptation step size.}
\end{figure}

\paragraph{Experiment $3$}
Now, we evaluate the performance of the proposed sparsity aware LMLS algorithms, i.e., $\ell_1$-CLMLS and $\ell_1$-WCLMLS in identifying a sparse system with variable sparsity. For this, similar to \cite{L1CNLMS}, we considered a randomly generated complex $30$th order filter. At first, the system is taken to be fully non-sparse, i.e., sparsity level is $0\%$. After one third of the time samples, the system is changed to moderately sparse system whose sparsity level is $50\%$. Finally, after two third of the time samples, the system is taken to be highly sparse with the associated sparsity level $90 \%$. The reference signal $d(n)$ is contaminated with the white Gaussian noise with variance $\sigma_{\vartheta}^{2}= 0.1$. The adaptation step size is fixed at $\mu=0.01$. The learning curves of $\ell_1$-CLMLS and $\ell_1$-WCLMLS along with $\ell_1$-CLMS and $\ell_1$-WCLMS are plotted in Fig.~\ref{L1CWLMS}. From Fig.~\ref{L1CWLMS}, it can be observed that the proposed $\ell_{1}$-WCLMLS has superior performance over the $\ell_{1}$-WCLMS.

%The simulation results presented in Fig.~\ref{L1CWLMS} reveals tha %superiority of the $\ell_1$-WCLMLS over the remaining algorithms.
\begin{figure}[h!]
\centering
 \subfigure[Sparse sytem identification \label{transient_NV}]{
\includegraphics[height=65mm,width=88mm]{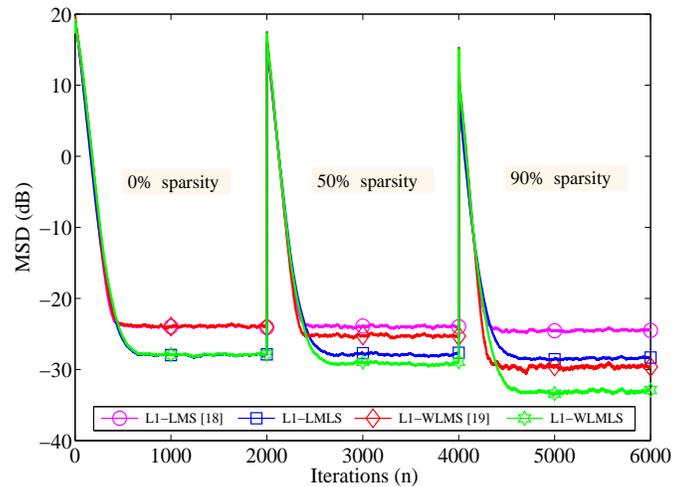} } \\
\subfigure[Compresible system identification \label{transient_Muw}]{
\includegraphics[height=65mm,width=88mm]{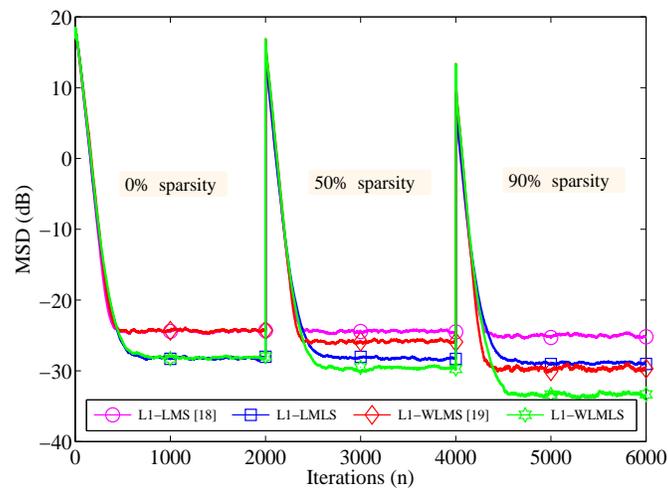} }
\caption{Learning curves of Sparsity-aware LMS (i.e.,  $\ell_1$-CLMS, $\ell_1$-WCLMS) and Sparsity-aware LMLS (i.e., $\ell_1$-CLMMS, $\ell_1$-WCLMLS) algorithms.}
\end{figure}

%\begin{figure}[t!]
%\centering
%\includegraphics [height=65mm,width=88mm]{L1CLMLS-f.eps}
%\caption{Learning curves of $\ell_1$-WCLMLS, $\ell_1$-CLMLS, $\ell_1$-WCLMS and $\ell_1$-CLMLS.}
%\label{L1CWLMS}
%\end{figure}

%\begin{figure}[h!]
%\centering
%\includegraphics [height=90mm,width=88mm]{CLMLS_SS_Theory_10_diff_NV.eps}
%\caption{Steady-state MSD of CLMLS Algorithm for White Input.}
%\label{steady-state}
%\end{figure}
                                                                                                                                                                                                                                                                                                                                                                                                             \section{Conclusions}
A novel linearly constrained adaptive filtering algorithm namely Constrained Least Mean Logarithmic Squares (CLMLS) is proposed. The proposed CLMLS exhibits improved performance over the existing CLMS algorithm. The mean-square performance of the proposed CLMLS is studied and validated in detail. The CLMLS is extended to sparse case by incorporating $\ell_1$-norm penalty into the CLMLS cost function. From the simulation results, it can be observed that the CLMLS/$\ell_{1}$-WLMLS can potentially replace CLMS/$\ell_{1}$-WLMS  in many practical applications that involve linearly constrained filtering problem. 
%\bibliography{mybibfile}

\bibliographystyle{IEEEtran}
%\bibliography{mybibfile}

\end{document}